\documentclass[conference]{IEEEtran}
\IEEEoverridecommandlockouts
\usepackage{cite}
\usepackage{amsmath,amssymb,amsfonts}
\usepackage{algorithmic}
\usepackage{graphicx}
\usepackage{textcomp}
\usepackage{xcolor}
\usepackage[utf8]{inputenc}

\usepackage{times}
\usepackage{epsfig}
\usepackage{paralist}
\usepackage{subfig}
\usepackage{comment}
\usepackage{graphics}
\usepackage{enumitem}
\usepackage{algorithm}
\usepackage{algorithmic}
\usepackage{multirow}
\usepackage{tabularx}
\usepackage{url}
\usepackage{wrapfig}
\usepackage{fancyvrb}
\usepackage{hyperref}
\usepackage{listings}
\def\BibTeX{{\rm B\kern-.05em{\sc i\kern-.025em b}\kern-.08em
    T\kern-.1667em\lower.7ex\hbox{E}\kern-.125emX}}

\newcommand{\pubtype}{~paper}

\newenvironment{packed_enum}{
\begin{enumerate}
  \setlength{\itemsep}{1pt}
  \setlength{\parskip}{0pt}
  \setlength{\parsep}{0pt}
}{\end{enumerate}}

\newenvironment{packed_itemize}{
\begin{itemize}
  \setlength{\itemsep}{1pt}
  \setlength{\parskip}{0pt}
  \setlength{\parsep}{0pt}
}{\end{itemize}}

\begin{document}

\title{\textbf{\texttt{MavVStream:}} Extending Database Capabilities for Situation Monitoring Using Extracted Video Contents \thanks{Dr. Sharma Chakravarthy was  partly supported by NSF Grants CCF-1955798 and CNS-2120393. Ms. Hafsa Billah was partially supported by NSF Grant CCF-1955798. Authors would like to thank Dr. Abhishek Santra for constructive suggestion for improving the paper.}}


\author{\IEEEauthorblockN{Hafsa Billah}
\IEEEauthorblockA{
\textit{IT Lab and CSE Department} \\
\textit{University of Texas at Arlington}\\
uxb7123@mavs.uta.edu}
\and
\IEEEauthorblockN{Mayur Arora}
\IEEEauthorblockA{
\textit{IT Lab and CSE Department} \\
\textit{University of Texas at Arlington}\\
mayur.arora@mavs.uta.edu}
\and
\IEEEauthorblockN{Sharma Chakravarthy}
\IEEEauthorblockA{
\textit{IT Lab and CSE Department} \\
\textit{University of Texas at Arlington}\\
sharmac@cse.uta.edu}}

\maketitle

\begin{abstract}
Query-based video situation detection (as opposed to manual or customized algorithms) is critical for diverse applications such as traffic monitoring, surveillance\footnote{Capital police surveillance footage from Ms. Pelosi's house was not being watched. If situations, such as ``break-in", ``shattering of glass", etc. could be expressed and monitored as continuous queries in real-time, alerts could have been raised in a timely manner!}, and other types of environmental/infrastructure monitoring.
 
Video contents are complex in terms of disparate object types and background information. Therefore, in addition to extracting complex contents using the latest vision technologies (including deep learning-based), their \textbf{representation as well as querying} pose different kinds of challenges. Once we have a representation to accommodate extracted contents, \textit{ad-hoc} querying on that will need new operators, along with their semantics and algorithms for their efficient computation. \textit{Extending database framework (representation and real-time querying) for processing queries on  \textbf{video contents \underline{extracted only once}} is critical and this effort is an initial step in that direction.}


In this paper, we extend the traditional relation to R++ (vector attributes) and arrables to accommodate video contents and extend CQL (Continuous Query Language) with a few new operators to query situations on the extended representation. Backward compatibility, ease-of-use, new operators (including spatial and temporal), and algorithms for efficient execution are discussed in this paper. Classes of queries are identified based on their complexity to evaluate with respect to video content. A large number of small and large video datasets have been used (some from the literature) to show how our work can be used on available datasets. Correctness of queries with manual ground truth, efficient evaluation as well as robustness of algorithms is demonstrated.~\textit{Our main contribution is couching a framework for a problem that is becoming very important as part of big data analytics based on a novel idea.}

\end{abstract}

\begin{IEEEkeywords}
Video content extraction, Continuous queries, Extended relational representation
\end{IEEEkeywords}

\section{Introduction}
\label{sec:motivation}
\noindent Ubiquity of inexpensive devices, such as dash/personal cams, mobile phones, and others has made it possible to generate large volumes of video data. Monitoring of these videos is important for postmortem analysis (or forensics), surveillance, security, or personal interest. \textit{Consider the Las Vegas shooting incident, where the suspect brought items through a door not being monitored. Automated monitoring of camera footage for specific situations, such as ``the same person entering multiple times within a specified time interval" or ``same person bringing bulky items multiple times during a specified time interval" would have generated a notification or alarm.}

Currently, these types of situations are monitored using custom solutions or manual analysis after the fact. Situation monitoring for \textbf{sensor Stream Processing (SP)} and \textbf{Complex Event Processing (CEP)} have matured. However, there is no database framework, representation, and query language for applying them to video contents (e.g., human, cars, etc.)

Although video streams have a fixed rate, the contents of each video frame are complex in terms of content extraction. Recently, more efficient \textbf{Video Content Extraction (VCE)} algorithms have been developed (e.g., YOLO~\cite{objectRecognintion/redmon2018yolov3}, M-RCNN~\cite{objectRecognintion/MRCNN}) to extract video information (e.g., feature vector, bounding box, background to some extent, activity, etc.). Without the appropriate representation of extracted contents, the traditional SP and CEP frameworks cannot be applied directly to extracted video contents as streams \cite{videoQuerying/yadav2021vidwin}. Although there exist some SP frameworks (e.g., Oracle Multimedia, Amazon Kinesis~\cite{AmazonKinesis}) for storing, searching, and retrieval of videos (or images) based on metadata, they have not \textit{pushed the boundaries to represent and query video contents continuously.}

In this paper, we postulate that video content analysis can be effectively enhanced by extending and synergistically integrating approaches from three domains (VCE, SP, and CEP) for obtaining an end-to-end holistic solution. VCE is employed for pre-processing each frame from a video to extract \textit{relevant contents}. The contents\footnote{Our focus is on the efficient processing  of the contents after extraction by the VCE algorithms. Real-time extraction of video contents is important, but a different problem and is being addressed by the video processing community and that is a prerequisite for video situation monitoring in real-time using our framework.} are streamed
to a continuous query processing system. The extracted contents need to be represented using expressive data models and processed with new operators to answer relevant queries for video situation monitoring.~\textit{A key advantage of this approach is that as each component advances, its effect can be leveraged by the other components to produce a system that is more than the sum of its parts.}

\subsection{Contributions}
The contributions of this paper are:
\begin{itemize}
\item A novel but well-defined and proven (low-risk) approach and its viability for video content analysis.
\item An expressive model (termed R++) for extracted information based on the relational data model.
\item Enhancing a relational model with multi-dimensional vector and arrable data types.
\item Extension of continuous query language with new operators/algorithms (e.g., join on feature vector, compression, direction, etc.) for video situation analysis (CQL-VA).
\item Extensive experimentation for accuracy, efficiency, robustness, and scalability.
\end{itemize}


\section{Challenges}
\label{sec:challenges}

\noindent The state-of-the-art VCE algorithms identify objects in each frame and track them across frames for a fixed amount of time (specified as a parameter) for labeling. The output of any VCE algorithm is dependent upon the lighting condition, camera position/angle, training data used, etc. Therefore, the robustness of the situations that can be identified from the extracted contents is dependent upon the accuracy and completeness of VCE algorithms. Our immediate goal is to extend CQL to formulate queries that can be posed and processed on the extracted video contents (\textbf{irrespective of the type of VCE algorithms employed}). The summary of situations that are addressed in this paper, the query types, and the computation types are shown in Table~\ref{tab:QueryType}.

\noindent \textbf{Problem Statement: } \textit{Use one or more state-of-the-art VCE algorithms for extracting contents of videos \underline{only once} and extend a proven database framework for representing the extracted contents using an expressive data model, and to support continuous queries (both \textit{ad-hoc} and ``what if") on the extracted contents. Following this, we expand the class of queries to more complex situations by adding relevant operators, their efficient processing, and associate accuracy (based on extraction accuracy) to the results obtained. The long-term goal is to create a data flow to automate real-time situation monitoring of videos.}

\begin{table}[!h]
\caption {\textmd{Query Types and Example Situations}}
\begin{tabular}{|m{1.5cm}|m{4.1cm}|m{1.6cm}|}
\hline
\textbf{Query Type} & \textbf{Example Situations} & \textbf{Computation Type} \\ \hline
Q1 (Search) & Presence of specific object classes (e.g., person) in a video, given the feature vector (or bounding box)  & Spatial \\
\hline
Q2 (Aggregation) & Counting the number of object classes/instances (e.g., person, car) in a given window. Can be used for extracting the duration of stay of an object, whether two objects entered within a time period, ... & Temporal  \\ \hline
Q3 (Window-based Join) & Same objects appearing in different videos (e.g., captured by entry or exit camera), individuals staying less than `n' minutes or leaving `n' minutes of each other, ... & Spatial \& Temporal  \\ \hline
Q4 (Direction) & Calculate net direction of movement of objects that are coming close, crossing, being exchanged, ... & Spatial \& Temporal \\
\hline
\end{tabular}
\label{tab:QueryType}
\end{table}

In our view, queries in Table~\ref{tab:QueryType} are starting points to video situation analysis involving aggregation/boolean queries where information from the video is aggregated along temporal or spatial, or both dimensions. For example, we have been able to successfully extend the class of queries in Table~\ref{tab:QueryType} to some of the following:  
\begin{packed_enum}
\item Are two people \textit{approaching} each other?
\item Did a person~\textit{remove} a delivered package? 
\item Are two people \textit{shaking hands or exchanging objects?}
\item \textit{Trajectory of a group} of people in a soccer game.
\end{packed_enum}

Note that the above situations are more complicated than the ones shown in Table~\ref{tab:QueryType} and require human pose identification (e.g., hand, arm, ...), background extraction (e.g., soccer field) as well as activity modeling (e.g., walking, bending, ...). Since a single VCE algorithm cannot extract all types of information, a combination of extraction algorithms is needed to build a richer extracted content representation. \textit{Hence, we focus on the queries in Table~\ref{tab:QueryType} in this paper as they are pre-requisites for detecting more complex situations shown above.}

Video analysis systems have been employing relational (e.g., VDBMS~\cite{videoQuerying/aref2003video}) or graph model~\cite{ vqa/xiong2019visual} for content representation and processing them without using stream processing approach. CQL, an extension of SQL (Structured Query Language), is used to query stream data types. However, \textit{video contents cannot be modeled with the existing data models, including arrables~\cite{lerner2003aquery} or array dbms~\cite{arrayDB/baumann2021array}}. For example, an object bounding box requires four different attributes to be represented using a relational model. 

Additionally, \textbf{video frames cannot be processed in any arbitrary order} since video context will be disrupted if frame sequence is broken. To compute Table~\ref{tab:QueryType} queries, consecutive frames where an object appears need to be brought together. Hence, an expressive data model to represent different types (e.g., vector, multi-dimensional arrays, categorical, ...) of information extracted by VCE algorithms is required for analysis. Besides, a set of operators are required to compare (e.g., matching feature vectors) and process (e.g., extracting direction) video contents. These operators should be backward compatible with the relational operators (e.g., select, aggregation, ...).

\begin{table*}[h]
\footnotesize
\caption {\textmd{Summary of existing video content analysis literature vs. MavVStream}}
\label{tab:qvc-related-work-comparison}
\vspace{5pt}
\centering
\begin{tabular}{|m{1.7cm}|m{3cm}|m{2cm}|m{5.5cm}|m{3.5cm}|}
\hline

\textbf{Category} & \textbf{System} & \textbf{Data Model} & \textbf{Supported} & \textbf{Not Supported}\\ 
\hline
\multirow{2}{1.9cm}{Custom solutions} & NoScope \cite{videoQuerying/kang2017noscope}, MIRIS \cite{videoQuerying/bastani2020miris}, SVQ++~\cite{videoQuerying/chao2020svq}
& Video & 
\multirow{2}{6.5cm}{Fixed set of queries (e.g., presence of object, car turning etc.) using deep learning} &  New class of queries (retraining required)\\ 
\cline{2-3} ~ & BLAZEIT \cite{videoQuerying/kang2018blazeit} & Relational & ~ & ~\\

\hline
\hline
Video streaming systems & Amazon Kinesis \cite{AmazonKinesis}, Oracle Multimedia  & Relational & Storage, search and retrieval of contents (or metadata) &  Querying (e.g., aggregation, join etc.) \\ \hline \hline

\multirow{2}{1.7cm}{Low-level content analysis frameworks} & VDBMS \cite{videoQuerying/aref2003video}  & \multirow{6}{*}{Relational}  & \textbf{Relational operators:} select, project, window based join on streams and non-streams & Aggregation, spatial and  temporal operators \\ \cline{4-5}
\cline{2-2}
& BilVideo \cite{videoQuerying/donderler2005bilvideo}, SVQL \cite{videoQuerying/svql2015}
& ~ & \textbf{Relational operators}: select, project, aggregation,
\textbf{Spatial operators:} bounding box overlap, direction and \textit{fixed event and video content database} & Any new type of video content or events, continuous query processing \\ \cline{4-5}
\cline{2-2}
& LVDBMS \cite{aved2014informatics} &  & \textbf{Low level operators}: appear, before, north etc. & 
 Window based joins\\ \cline{3-4}
\hline
\hline
\textbf{Extended CQL and Stream Processing} & \textbf{MavVStream (using R++, arrables, and CQL-VA)} & \textbf{R++, arrable} & 
\textbf{Relational operators (window based):}
select, project, join (nested \& hash), group by, aggregation

\textbf{CQL-VA operators:} compress consecutive tuples (CCT), consecutive join (cJoin)
with similarity matching condition for joining arrable, select on bounding box and arrable, direction 
& Background representation, intersect \\\hline
\end{tabular}
\vspace{-15pt}
\end{table*}

\textbf{Performing joins on video streams is also a challenge.} Traditional joins compute exact matches (by applying logical operators (e.g., $=$, $>$, etc.) over attributes that are not useful for comparing feature vectors, as \textit{they vary from frame to frame for the same object}. A matching condition (with an appropriate distance measure for comparing vectors) is required for this purpose, which is not supported in current systems.

Moreover, the number of frames in a video depends on video length, and frame rate, and the number of tuples generated for each frame is based on the number of objects present in that frame. Not all frames are necessary for many computations. As a result, operators are required for efficient processing without losing accuracy.
\textbf{Current SQL OVER clause cannot be used} to compute Q2 in Table~\ref{tab:QueryType}, as no time-based window is supported. A tuple-based window is inappropriate for processing this query since there can be multiple disjoint occurrences of the same object (need to be calculated separately). Q3 in Table~\ref{tab:QueryType} requires joining two different streams (from entry and exit cameras) and needs to check each object entering with all objects exiting with timestamp ts to ts+n as a window frame. Currently, OVER clause cannot specify this type of window frame. Therefore, it is important to have different types of windows (physical, logical~\cite{Babcock:PODS02:Models, DBLP:journals/vldb/ArasuBW06}, and predicate-based windows~\cite{DBLP:journals/sigmod/GhanemAE06}) introduced by CQL to the video analysis framework. Even other types of dynamic windows based on the presence of an object in consecutive frames are likely to be relevant.

\section{Relevant Work}
\label{sec:related-work} 

\noindent In Table~\ref{tab:qvc-related-work-comparison}, different categories of video situation analysis framework, their representation techniques and functionalities (both supported and not supported) are shown. 
The custom solutions shown in Table~\ref{tab:qvc-related-work-comparison}, are mostly deep learning approaches having fixed class of queries and the focus is on the optimization of neural network, speeding up model inference, etc.

We summarize the current solutions into three categories: (i) Video Streaming Systems, (ii) Low Level Content Analysis Frameworks, and (iii) Graph based Analysis. 

\noindent \textbf{Video Streaming Systems:} Many popular systems are available for stream data processing such as Apache Kafka~\cite{thein2014apache}, Apache Flink~\cite{Carbone:2017:SMA:3137765.3137777}, and Apache Heron~\cite{2015TwitterHS}. As mentioned earlier, these systems do not process video streams directly. Although the functionalities (see Table~\ref{tab:qvc-related-work-comparison}, row 2) of Amazon Kinesis \cite{AmazonKinesis}, SQL servers, and Oracle-Multimedia are useful for representing video content partially, representing the whole video context is still a problem.

\noindent \textbf{Low-level Content Analysis Frameworks:} 
These frameworks incorporate different operators and functionalities (see Table~\ref{tab:qvc-related-work-comparison}, row 3) for video analysis. 
Their major drawbacks are fixed content (or event) database (needs to be updated periodically) and limited continuous query processing support.

\noindent \textbf{Graph-based Analysis:} 
Recently, video analysis is also being explored using graph models~\cite{vqa/xiong2019visual,VideoRepresentation/yadav2020knowledge} to identify complex situations listed in Section~\ref{sec:challenges}. These solutions are either domain specific or involve custom operator set for each situation. Though VidCep~\cite{VideoRepresentation/yadav2020knowledge} supports window-based computations (e.g., selection, aggregation), none of the systems support situations involving joins and aggregation on different video streams.

\textbf{Differences with previous work:} 
Our work is significantly different from the custom solutions (including deep learning or graph neural networks) in terms of approach (e.g., machine learning vs query processing). These solutions require significant amount of training and retraining for identifying a class of activities (or situations). Additionally, the class of situations cannot be composed to identify complex situation sets described in Section~\ref{sec:challenges}.  
Our work differs from the low level content analysis frameworks in terms of the model and high level query language. These frameworks don't have support for continuous query processing, \textbf{joining multiple videos} and contains low level operators.
In this paper, we focus only on the relational model as the graph model is relatively new and comparison is difficult as the types and processing approaches of queries are different.



\section{Proposed Framework}
\label{sec:mavvstream-framework}
\begin{figure}[h]
\begin{center}
\includegraphics[width=0.5\textwidth, height=2.4in]{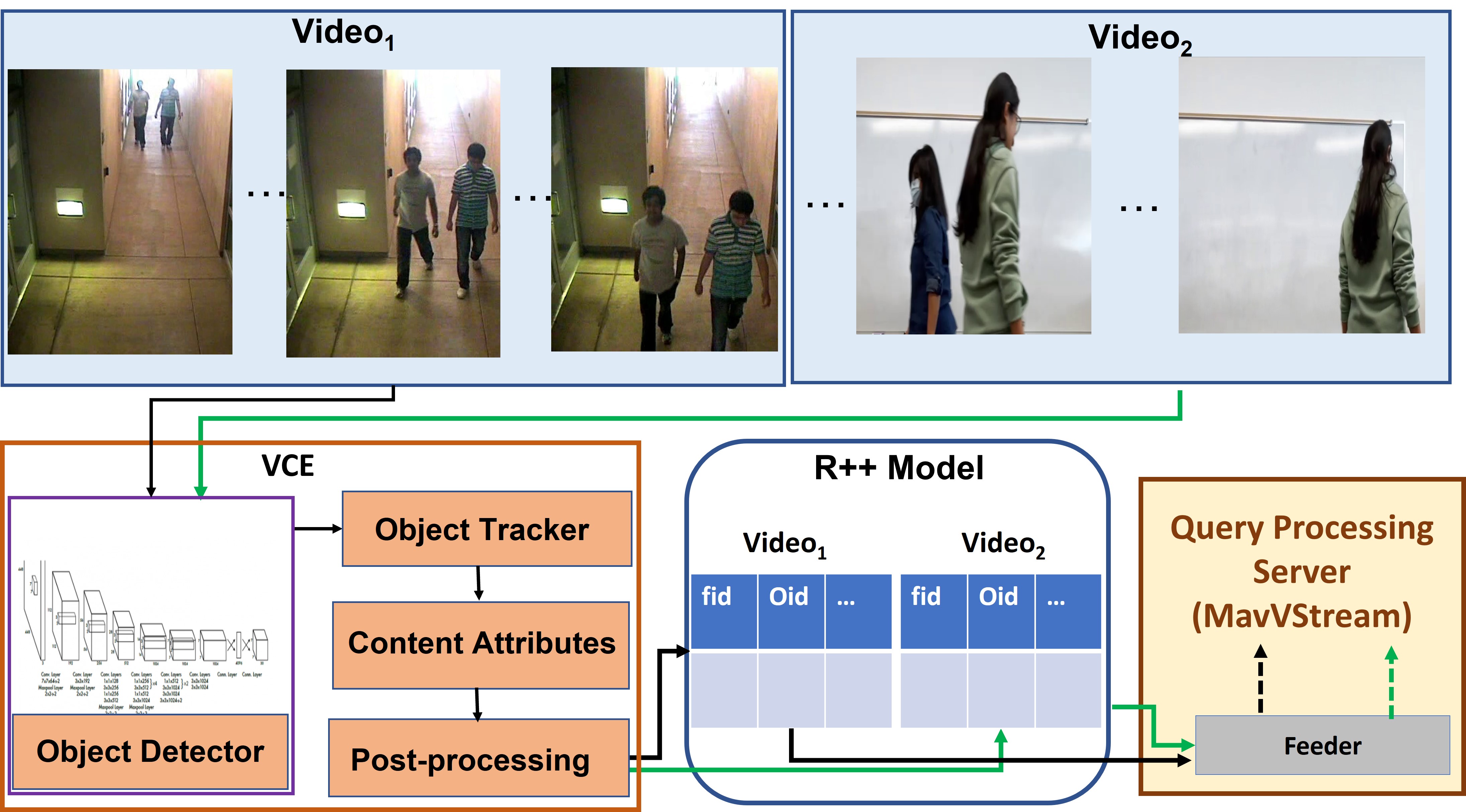}
\caption{\textbf{MavVStream Workflow:} Video Content Extraction including post-processing, Representation (using R++ model), and Continuous Query Processing (using MavVStream).}
\label{fig:mavvstream}
\end{center}
\vspace{-15pt}
\end{figure}

\noindent Our proposed video content analysis framework is composed of three modules: A) Video Content Extraction (VCE), B) modeling extracted contents using R++ and arrables, and C) processing continuous queries for situation detection. The system includes continuous query processing with new CQL-VA operators and their efficient evaluation, and the focus is on the accuracy and robustness of results. The components and data flow of the framework are shown in Fig.~\ref{fig:mavvstream}.  Note that, due to motion blur the image frames may appear blurry as image enhancement methods (e.g., motion correction) were not applied in any of the videos throughout the paper.

\subsection{Video Content Extraction (VCE)}
\label{sec:content-extraction}
\noindent A video is formed by capturing images at a specified rate (frames per second or fps) for a specific amount of time. The contents of each frame form a stream of static and moving objects and their spatial relationships over time. This stream is unbounded when used for surveillance monitoring. The purpose of VCE algorithms is to extract accurately (i.e., with high confidence) as much content from video frames as possible. This is dependent upon the environmental conditions (e.g., lighting, camera angle, movement, occlusion, ...) For example, VCE algorithms can misclassify when objects are partially present (or occluded). Besides, for different content types and purposes (e.g., activity modeling, human pose identification, scene parsing, ...), the category of VCE algorithms used differs\footnote{A multi-object tracking algorithm cannot be applied for human pose identification and vice versa.}. Hence, it is an important decision to determine what to extract based on the state-of-the-art VCE research as both \textit{the types of queries that can be posed and the correctness of answers} depend on the information extracted and their quality (or accuracy). 
The type of information extracted by the \textit{current} object detection and tracking algorithms is sufficient\footnote{Higher level complex queries require more information such as activity, pose, etc., which requires different VCE algorithms to be combined} to answer queries listed in Table~\ref{tab:QueryType}. Therefore, the VCE component of the proposed framework employs a state-of-the-art object detection algorithm (YOLO~\cite{objectRecognintion/redmon2018yolov3}) to extract moving objects, object types/classes, bounding boxes, and object feature vectors from each frame of the video. Individual objects are tracked (for assigning unique object id to each object across a maximum number of frames) using a state-of-the-art object tracking algorithm deep sort~\cite{Wojke2017simple}. Object tracking can pose some problems (e.g., two different objects can be tracked as a single object) depending on the accuracy of the object detection algorithm. Once all the content attributes are extracted, the raw outputs are post-processed for appropriate formatting and assigning a timestamp (ts) to each frame.

\textit{Since content extraction (object detection and tracking) is not the main goal} of this paper, we will assume that during the \textit{pre-processing phase}, the following content attributes can be \textbf{extracted as a minimum} from a video using any of the object detection and tracking algorithm available. 

\begin{enumerate}[label=\alph*)]
\item \textbf{Frame id (fid)}: Unique identifier of a frame
\item \textbf{Object id (oid)}: A unique identifier assigned to a particular object by the object tracking algorithm across video frames. If the video is long, some limits need to be imposed on the tracking duration.
\item \textbf{Object label}: Object type (e.g., car, person, etc.)
\item \textbf{Object bounding box ([BB])}: A rectangular area representing each object's location in a particular frame. It contains four elements: x and y coordinates of the lower left corner, width, and height.
\item \textbf{Object feature vector ([FV])}: A multi-dimensional vector representing features of an object. 
Ideally, a feature vector of an object should be similar across different video frames (or images). However, state-of-the-art VCE algorithms are unable to do so when proper environmental conditions are not present.

\end{enumerate}

It is worth noting that the quality of VCE algorithms is rapidly improving (in terms of accuracy and completeness of contents extracted). As a result, \textit{the object detection and tracking algorithms in this module can be replaced with new state-of-the-art ones (as and when publicly available).}

\subsection{R++ Model and Arrable}
\label{sec:representation}
\noindent Since, the traditional stream processing data models are not sufficient to represent the different content attributes extracted by VCE, we are\textit{enhancing the relational model (termed R++)} to include video contents listed above. R++ attributes can have traditional numerical values (e.g., fid, oid etc.), categorical values (e.g., object label, weekday, direction etc.), and vectors (e.g., multi-dimensional, ragged to capture BB and FV). 

The R++ table can also be converted to an arrable~\cite{lerner2003aquery} representation (using R2A operator of CQL-VA described in Section~\ref{sec:mavvstream-framework}) for ease of processing using simplified versions of queries.
When converted into an arrable by providing grouping and assuming order attributes it will have vector attributes, each of which can contain vector values indicated above due to R++ representation. In an arrable, all attributes that are vectors will have the same number of elements in them and are ordered by the ordering attributes. The R++ columns supporting vectors can contain vectors of any dimension and size. This allows support for any type and size of feature vectors (e.g., SIFT and YOLO generate feature vectors of size 128 and 512, respectively) or other information as they become available in the future.

\begin{figure}[h]
\begin{center}
\vspace{-6pt}
\includegraphics[width=0.48\textwidth, height=2.5 in]{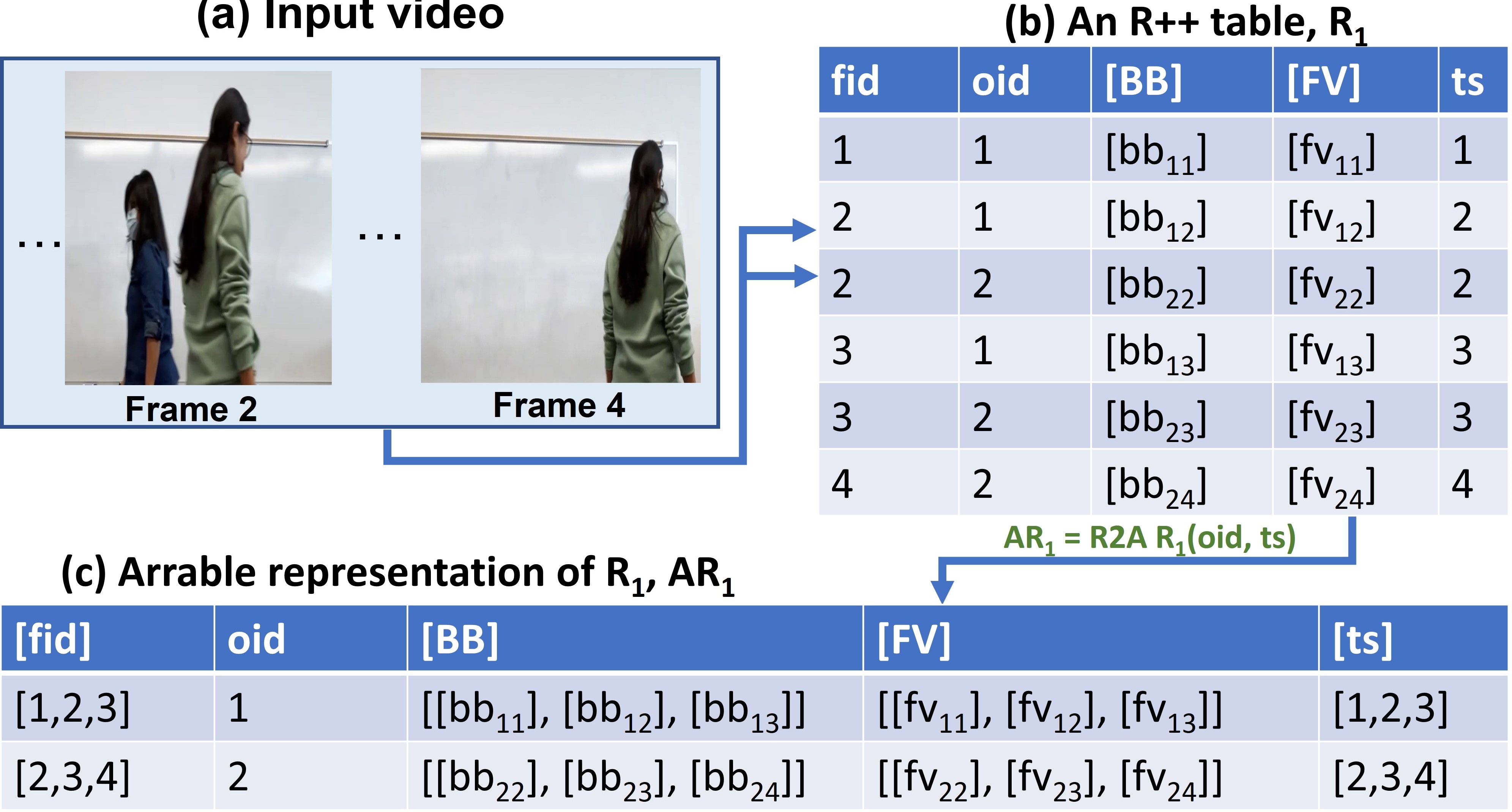}
\caption{\textmd{\textbf{Video Content Representation using R++ and Arrable}. $R_1$ shows the extracted contents using R++ (frame id as fid, object id as oid, bounding  box as  $[bb_{ik}]$, feature vector as $[fv_{ik}]$ of $i^{th}$ object in $k^{th}$ frame, and time stamp as ts. $R_1$ has been converted into an arrable $AR_1$ by grouping on oid and ordering on ts.}}
\label{fig:datamodelandoperator}
\end{center}
\vspace{-10pt}
\end{figure}

A tuple is created in an R++ table for each object detected in a video frame along with its associated attributes. An example of R++ table is shown in Fig.~\ref{fig:datamodelandoperator}(b). Here four different types of object attributes are shown. The attributes fid, oid, and ts are numerical types whereas [BB] and [FV] are vector types. The size of the [BB] vector is four. However, the size of the [FV] vector will vary depending on the feature vector type. In Fig.~\ref{fig:datamodelandoperator}, frame 2 contains two objects with object id 1 and 2. Therefore, there are two tuples for frame 2 (with different object ids) in the table. For supporting time-based windows, each frame is associated with an actual timestamp (shown as an integer for convenience). 

\textit{The difference between an R++ table and a traditional table (or relation in first normal form) is that not all operators can be applied on all columns.} For example, select operator has to be applied carefully on vector attributes. Although a bounding box can be selected using values and wild cards, it cannot be applied to a feature vector. Similarly, traditional join cannot be applied on a feature vector although it can be re-defined for a bounding box due to its finite size. The same is true for group by and aggregation operators. These have been implemented with checks in the MavVStream system so appropriate errors are given when used improperly. 

Applicability of relational operators and their backward compatibility on arrables has been established as part of the  Aquery~\cite{lerner2003aquery} semantics.  Arrable is a relation containing ordered collection of arrays/vectors, which means we can represent an ordered set of vectors associated with an attribute value with a single tuple. The different properties of computations (e.g., size preserving/not-preserving) on an arrable can be found in the Aquery~\cite{lerner2003aquery} paper. An example of arrable representation of an R++ table is shown in Fig.~\ref{fig:datamodelandoperator}(c). This table contains one tuple per oid, instead of containing a tuple for each fid and oid pair (see Fig.~\ref{fig:datamodelandoperator}(b)). This allows us to represent all the [BB] (or [FV], ts, fid, etc.) associated with an oid using a vector across the entire relation (or even a window which is a horizontal partitioning using ts).
It is possible to represent this table differently by grouping and ordering on other scalar attributes. Ordering attribute orders the elements of each column vector. 

\vspace{-5pt}
\subsection{Processing Continuous Queries for Situation Detection}

\noindent This MavVStream Framework component (shown in Fig.~\ref{fig:mavvstream}) is an extension of earlier stream processing system MavEStream\cite{Book/Chakravarthy09,DEBS/Chakravarthy08,ICDT/Jiang07} that had stream and event processing subsystems. MavEStream system has been  extended to include \textit{enhanced} relational representation, support for arrables, and new operators discussed in Section~\ref{sec:mavvstream-operators}. Continuous query processing differs from traditional DBMS query processing in many ways. First, query processing architecture is different as it is based on the stream input and its rate. Second, all computations are done in the main memory to avoid disk latency. Finally, as data is streaming, processing needs to be done without re-scanning the same data (i.e., one pass). For this, queues between operators and synopses are maintained as part of the query processing system. In addition, scheduling of operators and load shedding are important, if the input data rate is changing.
\begin{figure}[!h]
\begin{center}
\includegraphics[keepaspectratio=true,height=2.2in, width =0.5\textwidth]{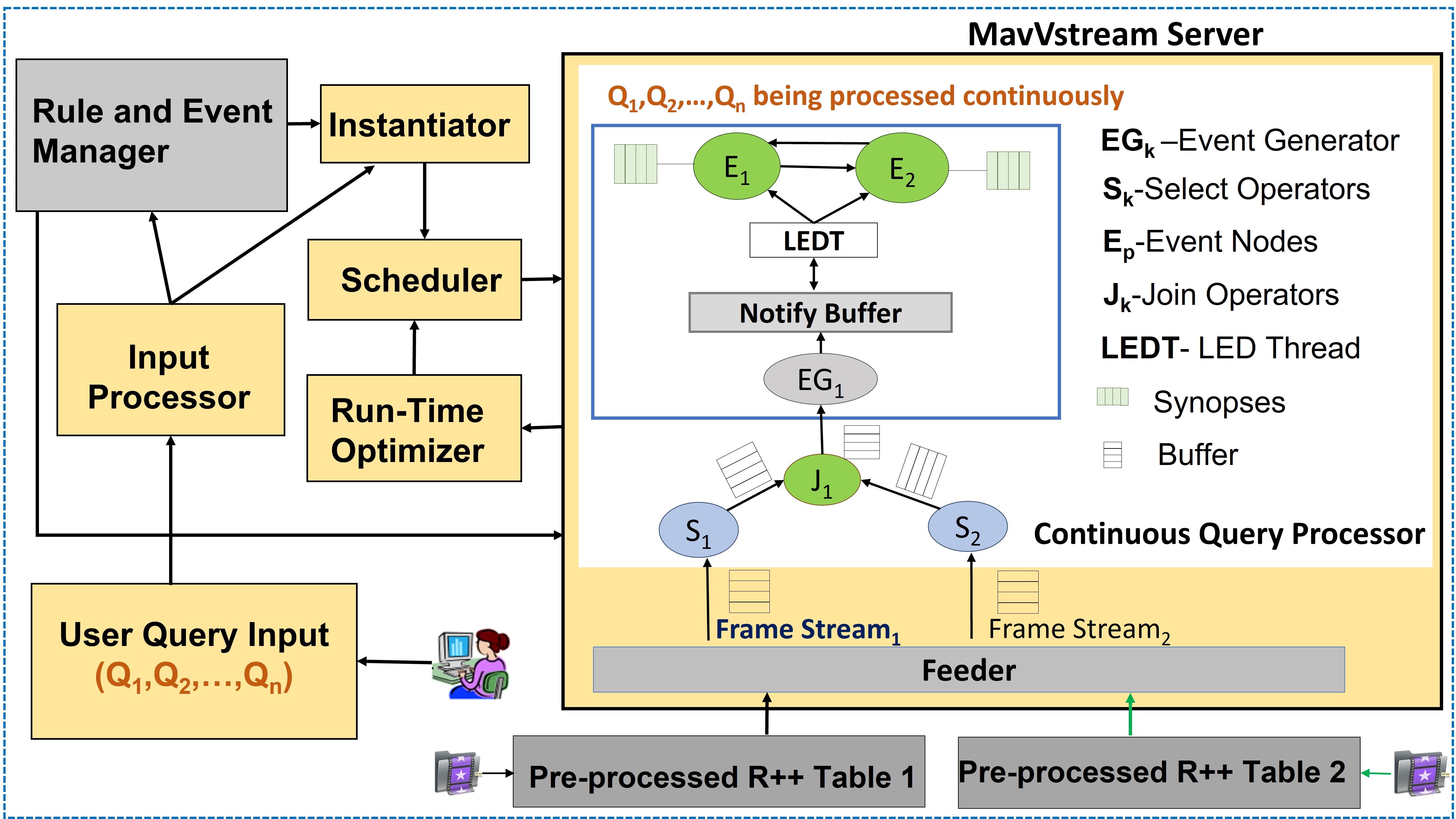}
\caption{MavVStream architecture and its modules~\cite{Book/Chakravarthy09,DEBS/Chakravarthy08,ICDT/Jiang07}}
\label{fig:mavvstream-details}
\end{center}
\end{figure}

The R++ model, arrables and some associated operators, and new operators (CQL-VA) are integrated into the video analysis architecture of MavVStream (Fig. \ref{fig:mavvstream-details}). 
This is a client-server architecture where users can submit CQL-VA queries to MavVStream server. Once the client submits a continuous query, a query plan is generated by the server. Each operator in the query plan object is associated with input and output buffers. Additionally, the underlying base of MavVStream, supports all basic relational operators along with aggregates as well as the CQL-VA operators discussed in Section~\ref{sec:mavvstream-operators}. The system also has support for both physical and logical windows with flexible window specifications with hop size to support disjoint and tumbling/rolling windows.

Operators and conditions to process vector attributes, such as bounding boxes and feature vectors have been added along with new arrable operators, such as compressing consecutive tuples (CCT), efficient join using consecutive Join (or cJoin), CCT with join using similarity match condition to compare objects for similarity. A multi-threaded feeder is used for feeding R++ tuples to the query processor. A round-robin scheduler schedules the operators and a run-time monitor is also available that can handle some quality-of-service (QoS) constraints. The complex event processing subsystem has also been integrated to provide primitive event detection capability (based on continuous queries) and compose them further.


\section{CQL-VA and the Relational Model}
\label{sec:mavvstream-operators}

\noindent CQL is compatible with the relational model. Similarly, arrable and operators proposed on Aquery~\cite{lerner2003aquery} have also been shown to be backward compatible with the relational model. CQL-VA introduced in this paper is also compatible with the relational model providing all the computations on that model. In this section, we discuss new operators, their semantics, need, and how they work with R++ and arrables. Operators that have been introduced are based on video processing requirements and queries in Table~\ref{tab:QueryType} and Section~\ref{sec:challenges}. These operators can also play a role in CQL-VA query optimization in the future. These operators process tuples of an R++ relation and arrables like any other relational operator, but have some constraints due to their applicability to specific types of attributes or data types. For example, in order to filter objects using the relational algebra SELECT ($\sigma$) operator, one needs to provide a feature vector or an image from which the feature vector can be extracted. Whereas, processing a bounding box attribute is relatively simpler using integer values and available wild cards. Joining video streams for object matching also requires dealing with feature vectors. These operations are not semantically valid for other attributes. This is important to understand\footnote{This is not any different from using certain operators based on attribute types. For example, average and sum are applicable only to numeric values.}. Care has been taken to introduce minimum number of primitive operators and composition is used (due to closure property) to express larger computations. 

The list of operators and conditions proposed as part of CQL-VA, their syntax, and complexity is shown in Table~\ref{tab:operator-syntax}.

\begin{table}[!h]
\centering
\caption { \textmd{\textbf{CQL-VA Operators and Conditions}. Here R$_i$: R++ relation; AR$_i$: arrable; gba, aoa: group by and assuming order attributes, respectively for creating arrables; N: number of tuples; M: number of objects in R++ or arrable relation (or window); $th$: threshold used in sMatch condition; $S$: complexity of sMatch; $a_i$: scalar attributes; G: number of rows (which are arrables); Nl$_{g}$, Nr$_{g}$: average number of tuples in left and right groups in join, respectively where $g=1,2,...,G$. dot notation used for attributes.}}
\begin{tabular}{|m{2.4cm}|m{3.37cm}|m{1.6cm}|}
\hline
\textbf{CQL-VA Operators} & \textbf{Syntax} & \textbf{Complexity} \\ \hline
Similarity Match (comparison operator) & \noindent\texttt{ sMatch\{(th)\} 
} & $\mathcal{O}(S)$\\
\hline

R++ to Arrable & \noindent\texttt{R2A (R$_1$, gba=R$_1$.a$_1$, aoa=R$_1$.a$_2$)} & $\mathcal{O}(N*\log N)$ \\ \hline
Compress Consecutive Tuples & \noindent\texttt{CCT  (AR$_1$, \{\underline{first}|last|both\})} & $\mathcal{O}(N)$ \\

\hline
Consecutive Join &
\noindent\texttt{ AR$_1$ cJoin (condition) AR$_2$  } &  $\mathcal{O}(G* Nl_{g} * Nr_{g} * S)$ \\ \hline
CCT Join &
\noindent \texttt{AR$_1$ cctJoin (condition)  AR$_2$} &  $\mathcal{O}(G^2 * S)$ \\ 
\hline
Direction &  \noindent\texttt{Direction (AR$_1$.[BB])} &  $\mathcal{O}(M)$ \\
\hline 
\end{tabular}
\label{tab:operator-syntax}
\end{table}
\noindent\textbf{Similarity Match Condition (\texttt{sMatch}):}
\label{sec:sMatch-condition}
Since feature vectors cannot be compared using the relational comparison operators, CQL-VA has added sMatch operator to compare different type of feature vectors. The feature vector size and semantics differ based upon the VCE algorithm used. A threshold can be optionally provided. Currently, two distance metrics cosine (for YOLO) and euclidean distance (for SIFT and histogram) are supported~\footnote{These distances are used by the computer vision research community to compare feature vectors.}. This condition can only be applied to the [FV] column type of any R++ table (or arrable). The output of the sMatch is the numerical distance between two feature vectors (ranging from 0-1). If the computed distance is greater than the given (or chosen) threshold ($th$), sMatch evaluates to True. Note that, the $th$ value given controls to what extent we can call two vectors similar and it is very sensitive to the environmental factors contributing to VCE algorithm accuracy mentioned earlier. Since, the complexity of the distance metrics is different, we quantify sMatch complexity by $\mathcal{O}(S)$. Obviously, sMatch is computationally more expensive than the relational comparison operators.

\noindent\textbf{R++ to Arrable Operator (\texttt{R2A}):}
The \texttt{R2A} operator converts an R++ table into an arrable representation. The group by (gba) and assuming order (aoa) parameters need to be \textit{scalar} attributes and perform grouping followed by ordering on each group according to AQuery semantics. \textit{This operator can be used, for example, to apply CCT join to improve efficiency. It can also be used to perform vector aggregate operations of AQuery, such as moving averages, again to improve efficiency.}

The group by operation using hashing takes $\mathcal{O}(N)$ time and sorting $G$ groups can be approximated by $\mathcal{O}(\log N)$ (upper bound) making the overall complexity $\mathcal{O}(N * \log N)$ for $N$ tuples in a R++ relation (or window).

\begin{figure}[!h]
\vspace{-10pt}
\begin{center}
\includegraphics[width=0.5\textwidth]{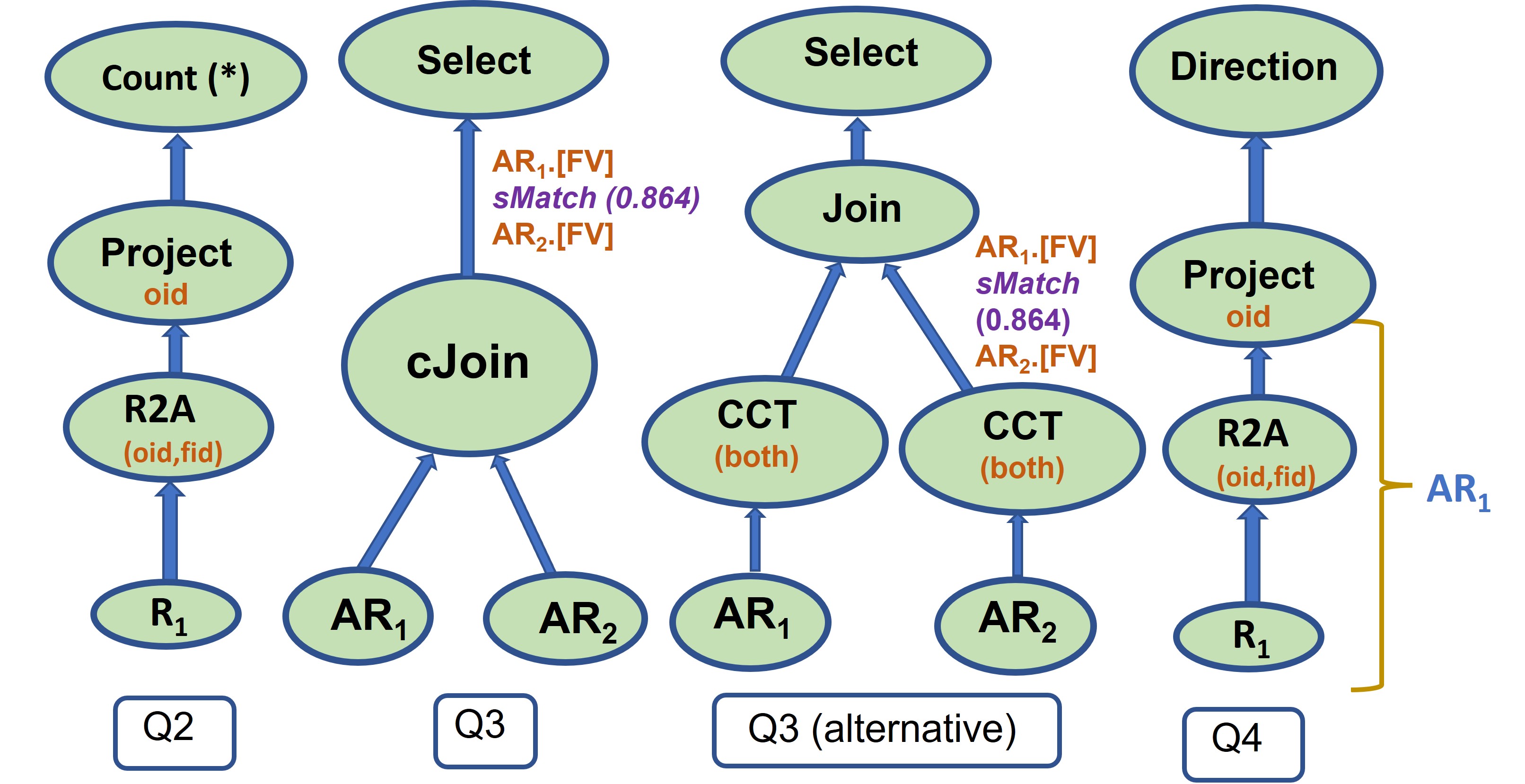}
\caption{\textmd{Operator Tree for Table~\ref{tab:QueryType} queries}. Alternative operator trees are shown for Q3. Here AR$_i$ is generated after applying \texttt{R2A} on R++ table R$_i$ as shown in Q4. Only sMatch conditions are shown due to limited space.}
\label{fig:operator tree}
\end{center}
\vspace{-15pt}
\end{figure}

\noindent \textbf{Compress Consecutive Tuples Operator (\texttt{CCT}):} A quirk of video frames is that the same object appears in multiple frames over the field of view. As each object is a tuple, computations such as select or join will compare the same object multiple times (quadratically for join). This operator reduces the occurrence of the same object multiple times by compressing the tuples generated by frames. Consecutive occurrences of the same object (whether as consecutive tuples in an R++ or arrable) are reduced to one or two occurrences (see queries Q2-Q4 in Table~\ref{tab:QueryType}). 

The CCT operator takes an arrable as input and compresses each vector attribute. It generates an R++ table or an arrable based on argument value, keeping the first/last item of each vector element or both. The relational operators such as selection (with sMatch condition and logical operators), join (on scalar attributes and sMatch condition) and aggregation are applicable to the result set generated by CCT operator (since they are applicable to arrable according to AQuery semantics). Fig.~\ref{fig:operator tree} shows query trees for Q2-Q4, where relational  operators are applied on top of CCT. Q3 is evaluated with and without the CCT operator to show the performance difference as shown in Section~\ref{sec:experimental-results}. 
Q2 in Table~\ref{tab:QueryType} only needs to access one tuple for an object id, whereas Q4 requires only two bounding box of an object (from the first and last frame). The DISTINCT operator from SQL does not have the property to remove \textit{consecutive multiple occurrences} from vector elements for \textit{efficiently} answering Q2-Q4.


\noindent \textbf{Consecutive Join Operator (\texttt{cJoin}:)} Although traditional join works on arrables using the added sMatch in addition to the relational comparison, it is still a join that compares each pair and in the case of videos, there are a lot of similar objects being compared.
cJoin was introduced as an option to not drop tuples (as in CCT) and improve accuracy.
This operator is applied after bringing the R++ into an arrable grouping on object id. Although cJoin can be applied on other attribute types, it provides the best efficiency for feature vector matches. Once a match is found, the rest of the tuples are not used (for the same object). In general, this will improve efficiency quite a bit although in the worst case,  sMatch may succeed only on the last tuple from both sides. cJoin will provide better accuracy than CCT join for the same computation as only the first/last or both are retained in CCT and they may not match. This has been validated experimentally in Section~\ref{sec:experimental-results}.

cJoin cannot be computed using a hash-based approach as the comparison is not equality. Hence, nested loop join is used for joins on feature vectors.  However, its performance is likely to be much better than traditional nested loop join. 
Assuming G groups in both (left: $l$, right: $r$) relations, each with $Nl_{g}$ and $Nr_{g}$ being average number of tuples in a group $g$, the worst case complexity will be $\sum_{g=1}^{G}\mathcal{O}(Nl_{g} * Nr_{g} * S)$. This can be simplified to  $\mathcal{O}(G*Nl_{g} * Nr_{g} * S)$. S is the cost of computing sMatch. The average complexity will be way better.


\noindent \textbf{CCT Join:} This operation first applies CCT operator on the input arrables, followed by traditional join with sMatch condition on the same attribute (see Fig.~\ref{fig:operator tree} Q4 alternative). This is typically applied on the feature vector attribute. Since CCT reduces the number of tuples in each row of the arrable to one or two, the number of comparisons made here is much less than cJoin and the accuracy will be less or equal to cJoin as well as traditional join.
Again, assuming G groups or rows in both arrables, each with $Nl_{g}$ and $Nr_{g}$ tuples in the group $g$, the worst case complexity will be $\mathcal{O}((G*2) * (G*2) * S)$. If first or last is chosen, it will be  $\mathcal{O}(G * G * S)$. The worst case complexity in simplified form is $\mathcal{O}(G^2 * S)$. As we indicated this is a time versus accuracy tradeoff.


\noindent \textbf{Direction Operator (\texttt{Direction}):} In videos, motion of objects is captured. If multiple objects are present in a video, one may want to identify whether they cross each other or groups of objects are moving in a single direction (e.g., in sports videos). In order to facilitate queries that require the direction of object movement (temporal queries like Q4), this primitive operator has been introduced. This operator can only be applied on the bounding box attribute of an arrable. 

This operator outputs one of the 8 directions as an enumerated type in a new column (used like an aggregate function) using the first and last bounding boxes. 
This operator can also be modified to compute object trajectory. The complexity of the direction operator is  $\mathcal{O}(M)$, where $M$ is the number of unique objects in that particular relation (or arrable) or window.



\section{Evaluation of CQL-VA Queries} 
\label{sec:cql-va-situation-detection}

\noindent Although, sMatch and direction are sufficient to process the query types indicated in Table~\ref{tab:QueryType}, we have introduced other operators for a reason. They are specific to video processing semantics and provide trade-offs with accuracy and response time as demonstrated in this paper. With these, optimization can be done by re-writing queries, where meaningful to take advantage of it.
Here we discuss different aspects of CQL-VA evaluation (accuracy, efficiency, robustness) for Table~\ref{tab:QueryType} queries. 

\subsection{Query Formulation}
\noindent\textbf{Q1 (Search)}: This query is formulated on an R++ table using sMatch condition in the where clause.\\

\noindent\textbf{Q2 (Count):} \textit{Count distinct persons in the video}

\begin{Verbatim}[commandchars=\\\{\}]
Select count(*)
From
    (Select AR\(\sb{1}\).oid 
    From 
    (R2A (R\(\sb{1}\), R\(\sb{1}\).oid, R\(\sb{1}\).fid))  AR\(\sb{1}\)
    Where (R\(\sb{1}\).label = "person"))
\end{Verbatim}

\noindent\textbf{Q3 (Join):} Identify same person in two videos
\begin{Verbatim}[commandchars=\\\{\}]
Select AR\(\sb{1}\).oid, AR\(\sb{2}\).oid
From 
    (R2A (R\(\sb{1}\), R\(\sb{1}\).oid, R\(\sb{1}\).fid))  AR\(\sb{1}\)
    cJoin
    (R2A (R\(\sb{2}\), R\(\sb{2}\).oid, R\(\sb{2}\).fid)) AR\(\sb{2}\)
    on AR\(\sb{1}\).[FV] sMatch(0.008) AR\(\sb{2}\).[FV]
\end{Verbatim}

There are alternative ways to formulate this query (see Fig.~\ref{fig:operator tree} Q3 alternative). Additionally, a regular join with sMatch condition can also be used.

\noindent\textbf{Q4 (Direction):} \textit{Output direction of motion of objects.} Query formulation is similar to Q2. Instead of the count, \texttt{Direction} operator is applied in the select clause (see Fig.~\ref{fig:operator tree}).
\\

\vspace{-15pt}
\subsection{Query Evaluation and Accuracy}
\noindent VCE algorithms do not correctly identify all the objects present in a video correctly (specifically if the video is highly dissimilar from the training videos and due to environmental conditions). The ground truth (GT) of an original video doesn't match all the time with the VCE output in terms of object presence. As a result, two different objects can be assigned same oid by VCE as shown in Fig.~\ref{fig:query2-output} (actual case). Therefore, query accuracy is evaluated for video and VCE output separately as Acc(v) and Acc(vce), respectively throughout this paper. \textit{Our goal is to match VCE ground truth as we are not improving the accuracy of VCE algorithms}. The standard accuracy formula shown below is used.

\begin{equation}
\small
    Accuracy = \frac{TP+TN}{TP+TN+FP+FN}
\end{equation}

\subsection{Query Efficiency and Scalability}
\noindent Efficiency is the time taken for the same computation using different operators preserving accuracy as closely as possible.
Scalability measures the performance of the operators when the video size increases. For scalability, we have used videos of different lengths (maximum 44.5 min.) and computed the time taken. Since there is a scarcity of large videos containing all the situations of interest in this paper, we have trimmed and merged videos from the CAMNET~\cite{DataSet/Camnet} dataset to increase video length.

Additionally, the performance of the different types of joins in CQL-VA is compared for different video lengths to illustrate the efficiency of introduced operators.
\subsection{Robustness} 
\noindent Typically, queries detect a specific situation in a video. Multiple queries are used to detect different situations on the \textbf{same} video. The robustness of a query is high if \textit{it does not detect noise or other situations} in the video. For this validation, we have taken videos used for Q3 (which is a join and the query is more complex) and added (concatenated) other video parts (from CAMNET~\cite{DataSet/Camnet} dataset) which do not have the situation being detected by Q3. If we get the same accuracy, then we can say that our query formulation and computation are robust. This establishes that a situation can be detected correctly even in the presence of other arbitrary situations. This is shown in Section~\ref{sec:experimental-results}.


\section{Experimental Results}
\label{sec:experimental-results}
\noindent In this work, we have used two different datasets: CAMNET~\cite{DataSet/Camnet} and MavVid~\cite{DataSet/MavVid} for testing all the situations listed in Table~\ref{tab:QueryType} which includes videos of people passing by a place (e.g., hallway) in complex lighting conditions (e.g., shadow, reflection, etc.) and background. The description of datasets is shown in Table~\ref{tab:dataSetDescription}. Note that, the number of tuples generated by a video can be different from the number of frames (which determines  the length of the video) because of empty frames and the number of objects present in each frame. For example, Video id (V$_{id}$) 1 in CAMNET is larger in length than V$_{id}$ 23 but V$_{id}$ 23 has more tuples. \textit{Processing time for a video is determined by the number of tuples generated by the VCE more than the video length.} All the videos in the dataset were pre-processed using the YOLO implementation available in ~\cite{Implementation/YOLO} using an NVIDIA Quadro RTX 5000 GPU with 10 GB memory. The query processing experiments were conducted on the same machine with 2 processors (Intel Xeon), 48 cores, and 746 GB of main memory.

The situations of interest in the existing video stream processing systems shown in Table~\ref{tab:qvc-related-work-comparison}  are different than the situations addressed in this paper. Though VidCEP~\cite{VideoRepresentation/yadav2020knowledge} have investigated a similar query like Q1 in Table~\ref{tab:QueryType}, they counted objects from each video frame (e.g., counting the same object multiple times over a period) on 10 videos of DETRAC~\cite{Dataset/CVIU_UA-DETRAC} dataset. In their query, they did not specify which 10 videos of the dataset were used. In addition, Q2-Q4 have not been addressed in any of the video stream processing literature. Thus, comparing the results of the proposed solution with existing ones is extremely difficult. It's worth mentioning that, the video analysis frameworks (\cite{VideoRepresentation/yadav2020knowledge, vqa/xiong2019visual}) have not also compared their results due to the above reasons.

\begin{table}[ht]
\caption {\textmd{Dataset Description}}
\begin{tabular}{|m{1.2cm}|m{1.8cm}|m{2cm}|m{0.8cm}|m{0.6cm}|}
\hline
\textbf{Dataset}  & \textbf{Video Id (V$_{id}$)} &  \textbf{Length (in minutes \& seconds)} & \textbf{Frames} &  \textbf{Tuples}\\ \hline

\multirow{2}{0.8cm}{MavVid \cite{DataSet/MavVid}} &  3439 & 44s & 1139 & 173 \\
\cline{2-5}
 ~ & 3437 & 40s & 832 & 3200 \\
\cline{2-5}
 ~ & 3441 & 44s & 898 & 2437 \\\cline{2-5}
 ~ & 3443 & 60s & 1136 & 362 \\\cline{2-5}
 ~ & 3453 & 50s & 1017 & 1004 \\\cline{2-5}
 ~ & 3457 & 44s & 893 & 1643 \\\cline{2-5}
\hline
\multicolumn{2}{|c|}{Total} & 3 min. 18s & 6270& 8819 \\
\hline
\hline
\multirow{2}{1cm}{CAMNET \cite{DataSet/Camnet}} & 1 & 23 min. 57s & 26338 & 11691 \\\cline{2-5}
 ~ & 23 & 20 min. 30s & 24876 & 15700 \\\cline{2-5}
 ~ & 29 & 22 min. & 26078 & 11035 \\\hline
\multicolumn{2}{|c|}{Total} & 66 min. 27s & 77292 & 38426 \\
\hline
\end{tabular}
\label{tab:dataSetDescription}
\end{table}

\begin{table}
\centering
\caption { \textmd{\textbf{Accuracy of Query Q1}. $W_{t}$: total number of windows, $W_{s}$: window size in seconds.}}
\begin{tabular}{|l|l|l|l|p{01cm}|l|l|}
\hline
\textbf{Dataset} & \textbf{$V_{id}$} & \textbf{Acc(v)} & \textbf{Acc(vce)} & \textbf{Best \textit{th}} for accuracy & W$_s$ & W$_t$\\
\hline
\multirow{2}{1cm}{MavVid \cite{DataSet/MavVid}} & 3441 &  100\% & 100\% & 0.8500 & 120 & 1 \\ \cline{2-7}
~& 3443 &  100\% & 100\% & 0.8640 & 120 & 1 \\ \cline{2-7}
~& 3453  & 100\% & 100\% &  0.835 & 120 & 1 \\ \hline
\end{tabular}
\label{tab:Query1-Result}
\end{table}

\subsection{Accuracy}
\label{sec:result}
\noindent\textbf{Q1 (Searching):}
The experiments for Q1 were conducted on MavVid Dataset (shown in Table~\ref{tab:Query1-Result}) using a time-based disjoint window of 120s. The videos were searched for the same object (using a feature vector) by changing \textit{th} values. The best \textit{th} value represents the threshold for which we obtain the highest accuracy. A small deviation from this value affects the accuracy significantly as feature vectors vary significantly for the same object across frames. For example, changing the \textit{th} values of the video 3441 in Table~\ref{tab:Query1-Result}) from 0.85 to 0.7990 results in 0\% accuracy. This is true for other videos as well and represents the sensitivity of the th values given to sMatch. \textbf{Determining \textit{th} value automatically is an open problem.}

\noindent\textbf{Q2 (Aggregation) and Q4 (Direction):}
The experiments for Q2 and Q4 were performed on both the CAMNET and MavVid datasets using the query formulation shown earlier (see Fig.~\ref{fig:operator tree} for query tree and Table~\ref{tab:Query2-Result} for results) with a time-based disjoint window of 1000s and 120s, respectively. Since MavVid videos are much smaller than CAMNET, the whole video was a window.

\begin{table}[htb]
\centering
\caption {\textmd{\textbf{Accuracy of Query Q2 and Q4 on different datasets}. $W_{t}$: total number of windows, $W_{s}$: window size in seconds. Threshold is not required for this as sMatch not used.}
}
\begin{tabular}{|p{1.8cm}|p{0.5cm}|p{0.6cm}|l|l|l|l|}
\hline
\textbf{Dataset}  & \textbf{$V_{id}$} & \textbf{Query} & \textbf{Acc(v)} & \textbf{Acc(vce)} &  \textbf{$W_s$} & \textbf{$W_{t}$}\\\hline

\multirow{2}{1.8cm}{CAMNET ~\cite{DataSet/Camnet}} & \multirow{2}{1cm}{1} & Q2 &  89\% & 100\%   & 1000 & 7\\\cline{3-7}
~& ~ & Q4 & 77\% & 100\% & 1000 &7 \\\cline{2-7}
~ & 23 & Q2 & 70\% &  100\% & 1000 & 10 \\ \cline{3-7}
~ & ~ & Q4 & 71\% & 100\% & 1000 & 10 \\\cline{2-7}
~ & 29 & Q2 & 60\% &  100\% & 1000 & 8 \\ \cline{3-7}
~ & ~ & Q4 & 75\% & 100\% & 1000 & 8 \\\cline{2-7}
\hline \hline

\multirow{2}{1.8cm}{MavVid~\cite{DataSet/MavVid}} & \multirow{2}{1cm}{3441} & Q2 &  66.67\% & 100\%    & 120 & 1\\\cline{3-7}
~& ~ & Q4 & 66.67\% & 100\% & 120 &1 \\\cline{2-7}
~ & 3443 & Q2 & 66.67\% &  100\% & 120 & 1 \\ \cline{3-7}
~ & ~ & Q4 & 66.67\% & 100\% & 120 & 1 \\\cline{2-7}
~ & 3453 & Q2 & 66.67\% &  100\% & 120 & 1 \\ \cline{3-7}
~ & ~ & Q4 & 66.67\% & 100\% & 120 & 1 \\\cline{2-7}

\hline 


\end{tabular}
\label{tab:Query2-Result}
\end{table}

\begin{figure}
\begin{center}
\includegraphics[keepaspectratio=true, width=0.48\textwidth, height = 0.45\textwidth]{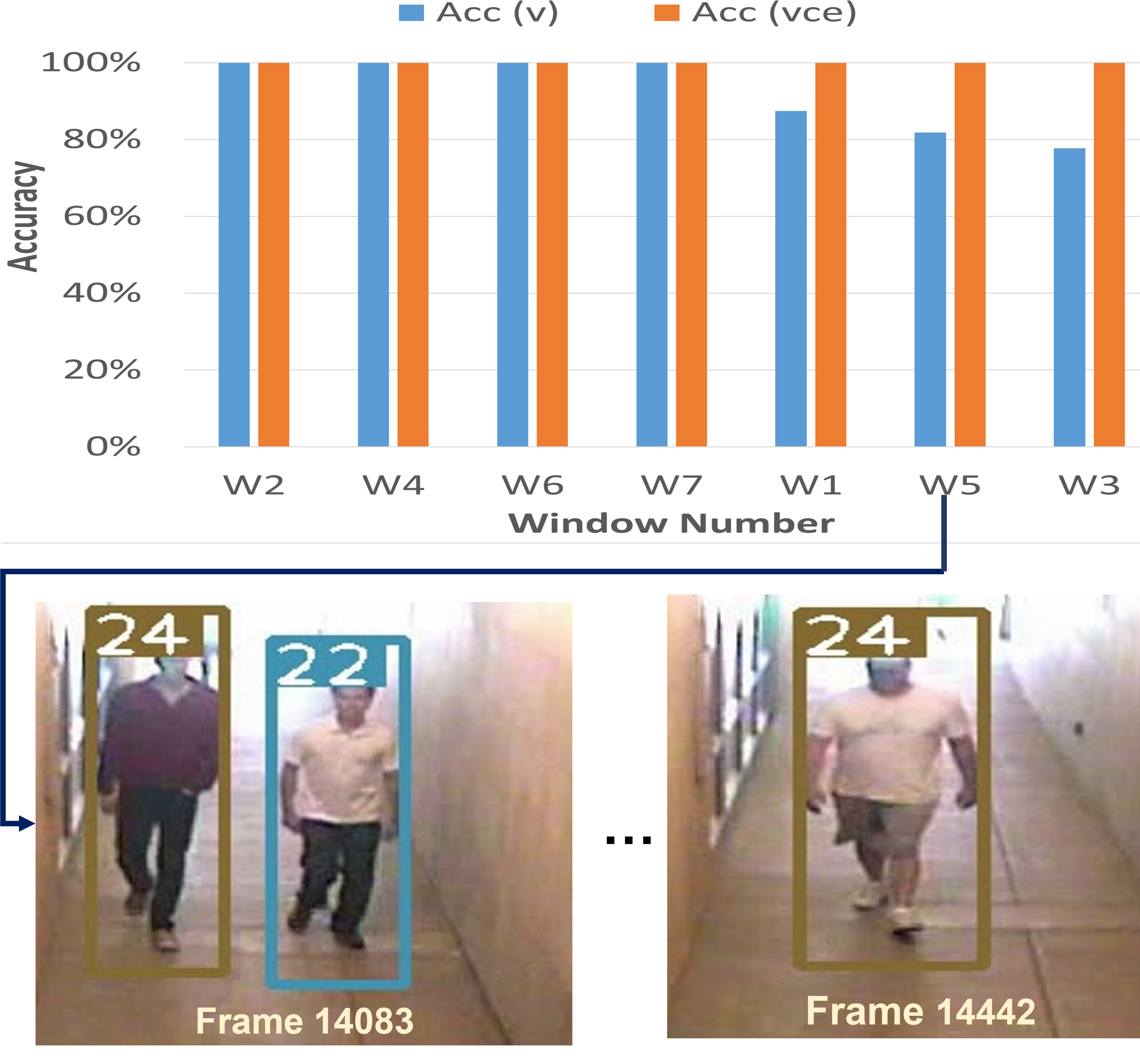}
\caption{\textmd{Q2: Accuracy (Acc(v) and Acc(vce)) of each window for $V_{id} = 1$ in CAMNET~\cite{DataSet/Camnet} dataset is shown here. W$_i$ represents i$^{th}$ window. In window 5, the two different objects are assigned the same oid (24), shown in top left corner of the rectangles in each image. The image frames extracted from CAMNET~\cite{DataSet/Camnet} dataset appear blurry because of the camera used to capture the videos.}}
\label{fig:query2-output}
\end{center}
\end{figure}


\begin{table*}[ht]
\centering

\caption { \textmd{\textbf{Accuracy of Query Q3 on different video pairs on MavVid dataset. $V_{id_i}$}: video id, $O_i$: unique identifier to represent an object across videos (this is not oid), GT(vce): ground truth for VCE output.}}
\begin{tabular}{|m{0.8cm}|l|m{1cm}|m{1cm}|m{2cm}|m{1cm}|m{1cm}|m{1cm}|m{4.6cm}|}
\hline
\textbf{Dataset} &  \textbf{$V_{id_1} \bowtie V_{id_2}$}  & \textbf{$V_{id_1}$ objects}  & \textbf{$V_{id_2}$ objects} & \textbf{GT(vce)} & \textbf{Join Acc(vce)} & \textbf{cJoin Acc(vce)} & \textbf{CCT Join Acc(vce)} & Remarks\\ \hline

\multirow{2}{1 cm}{MavVid
\cite{DataSet/MavVid}} & $3437 \bowtie 3437$ & O$_1$,O$_2$, O$_3$,O$_5$ & O$_1$,O$_2$, O$_3$,O$_5$	& (O$_1$,O$_1$),(O$_2$,O$_2$),
(O$_3$,O$_3$),(O$_5$,O$_5$)  & 100\% & 100\% & 100\% & \textbf{Self join}\\ \cline{2-9}

 & $3437 \bowtie 3439$ & O$_1$,O$_2$, O$_3$,O$_5$ & O$_1$,O$_3$ 	& (O$_1$,O$_1$),(O$_3$,O$_3$)  &  62.5\% & 62.5\% &  62.5\% & More TP and FP in join than cJoin and CCT join. O$_1$ detected as person in 15 frames only and door in rest.\\ \cline{2-9}

 & $3437 \bowtie 3457$ & O$_1$,O$_2$, O$_3$,O$_4$ & O$_1$,O$_6$, O$_7$,O$_8$ & (O$_1$,O$_1$)  & 93.8\% &  75\% &  75\% & O$_1$ has different color clothes.  \\ \cline{2-9}
 \hline
\end{tabular}
\label{tab:Query4-Result}
\qquad
\end{table*}

Here, Acc(vce) is 100\% for both queries in all the input videos. Acc(v) drops for these queries as VCE does not produce the correct output. VCE output is shown for two video frames in Fig.~\ref{fig:query2-output}. As can be seen, in  window 5 different objects (see frames 14083 and 14442) are identified as a single object and given the same oid, which affects accuracy. This is true for Q2 as well, as it will compute one net direction for two different objects.

\noindent\textbf{Q3 (Join) common objects from two videos:}
The experiments for Join were conducted on three videos of MavVid dataset (shown in Table~\ref{tab:Query4-Result}) using different join operations (regular join with sMatch condition, cJoin and CCT Join also with sMatch and using the same \textit{th} value) of CQL-VA. For CCT Join \textit{both} was chosen as a parameter. Here, O$_i$ is used to represent objects uniformly across videos. GT(vce) shows the ground truth with respect to VCE along with the common object pairs that should be generated by join. All the join operations have the same accuracy (different TP and FP) when videos 3437 and 3439 are joined. Although regular join extracts all the object pairs correctly from ground truth, it brings out more FP as compared to cJoin and CCT Join. CCT Join gives the same accuracy as cJoin since the feature vectors in the first frame were matched from both videos. Also, object $O_1$ in video 3437 is detected as a person only in 15 frames and a door in the rest of the frames. This depicts that the feature vectors also contain the properties of door, whereas $O_1$ in 3439 does not have such background. Therefore, finding a match for $O_1$ from both videos even with the best threshold value is difficult. 

On the other hand, in video 3457 object, $O_1$ is wearing different colored clothes. Hence, the feature vectors are very dissimilar for the same object $O_1$ in 3437. Regular join was able to identify the TP but it has higher FP and TN than cJoin and CCT join. On the other hand, cJoin and CCT join brings out no TP in this case. 

We have also performed \textbf{self join} of video 3437 to validate the correctness of the operator implementations. For self-join, we get 100\% accuracy with respect to VCE as expected for all flavors of join. It also shows, if feature vectors are consistent, then the results will be more accurate.

\subsection{Efficiency and Scalability}
\begin{figure}[ht]
\vspace{2pt}
\begin{center}
\includegraphics[width=0.5\textwidth, height = 0.3\textwidth]{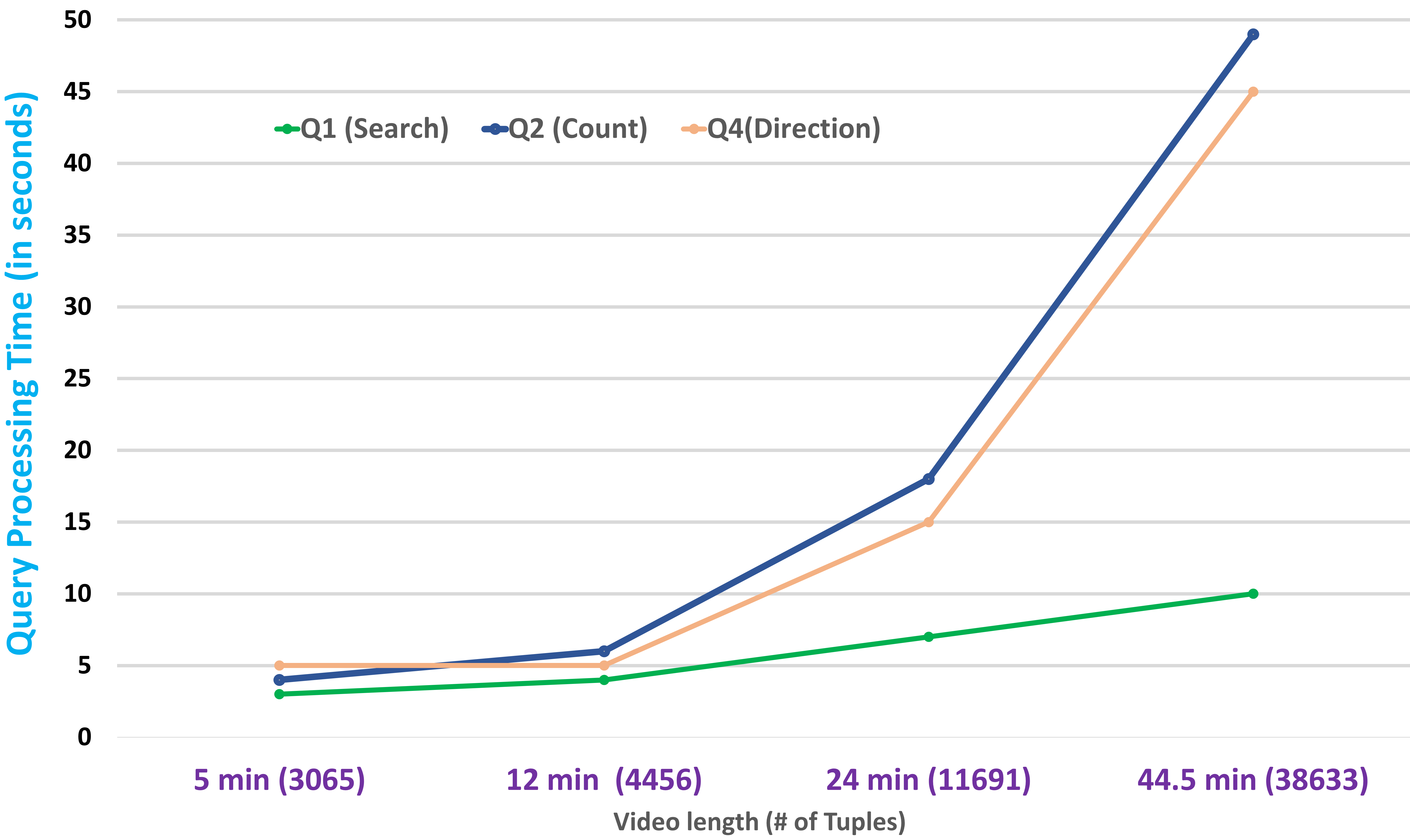}
\caption{\textmd{ Query processing time for different length videos from CAMNET dataset. A time-based disjoint window of 1000s was applied here.  For Q1, a threshold value of 0.95 was used.}}
\label{fig:query-efficiency}
\end{center}
\end{figure}

\noindent Efficiency experiments were performed on Video 1 and 23 of CAMNET dataset. We have created 5 and 12 minutes videos from Video 1 and 23 by taking the first 5 and 12 minutes from each video, and a 44.5 minutes video by merging videos 1 and 23. The purpose of this was to increase the size of the videos 2 times to understand the processing time taken by the queries. In Fig.~\ref{fig:query-efficiency}, the performance of Q1, Q2, and Q4 is shown. With increased video size, the query processing time increases linearly for Q1, Q2,  and Q4. The values of the curve vary based on the operator used and the time taken for its computation. Since Q1 is a select operation, it takes the least amount of time among all the queries.

\begin{figure}[htb]
\begin{center}
\includegraphics[width=0.5\textwidth, height = 0.3\textwidth]{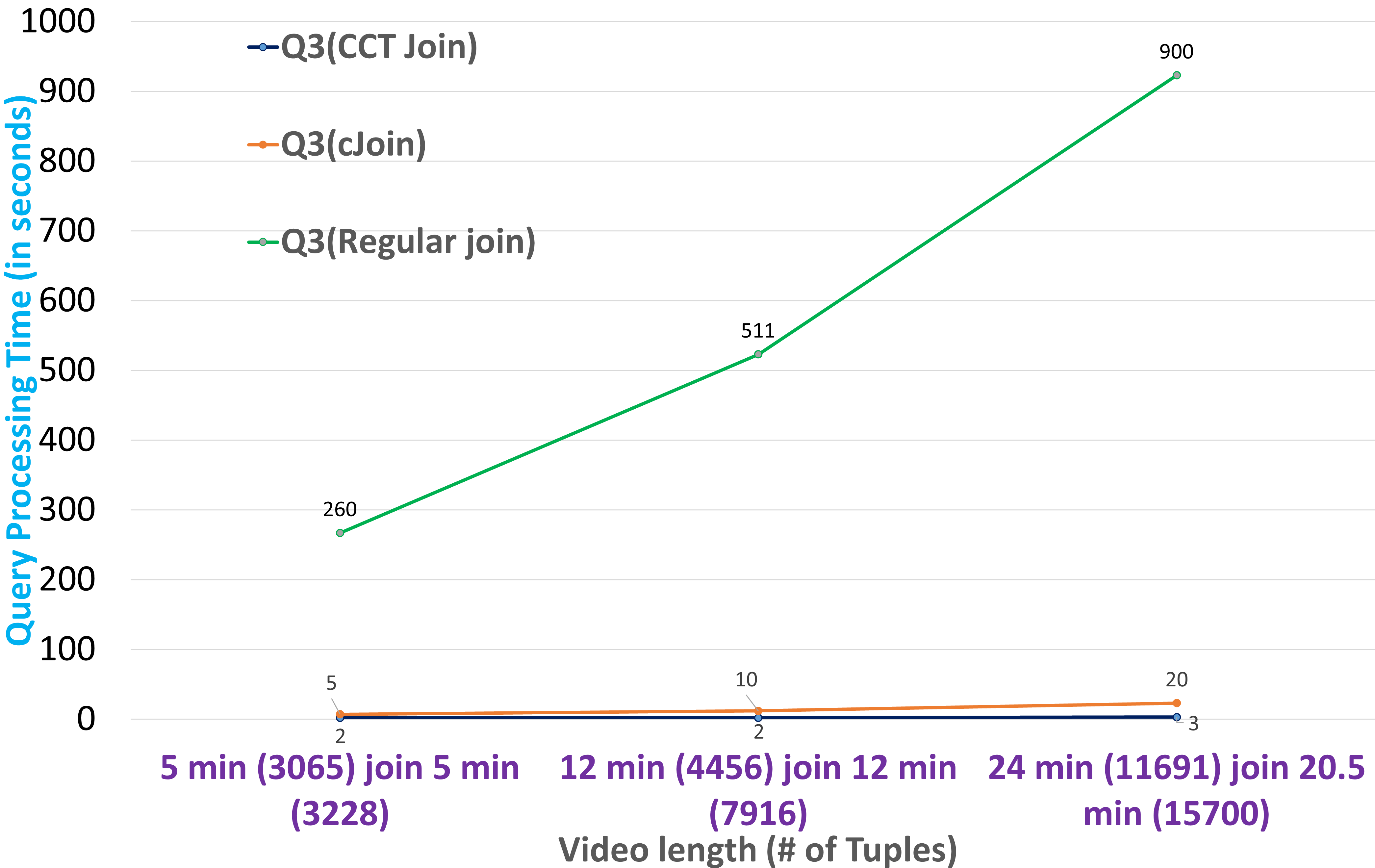}
\caption{\textmd{ Q3 processing time for different length videos from CAMNET dataset. Time based window size of 10000s used.}}
\label{fig:query-efficiency-Q3}
\end{center}
\vspace{-5pt}
\end{figure}

The performance of Q3 was measured in Fig.~\ref{fig:query-efficiency-Q3} by joining the same size videos sampled from 1 and 23 as well as joining 1 and 23. As we mentioned earlier, the regular join time (max 900s) is significantly higher than cJoin and CCT join, since the number of comparisons made is higher. This validates our intuition for introducing these operators. Also, the accuracy varied between 7\% and 15\% for different joins also as expected. \textbf{New CQL-VA operators introduced perform more than an order of magnitude faster as compared to the traditional join without sacrificing accuracy very much!}

\subsection{Robustness}
\noindent The robustness experiment was performed by adding arbitrary video at the end of video 3437 and 3439 (represented as 3437$^{+}$ and 3439$^{+}$) from MavVid dataset. The results are shown in Table~\ref{tab:Query3-robustness}. Since the additional videos merged do not contain any common objects, the ground truth remains the same. However, the accuracy improves from 62.5\% to 80\% (for all the join operators) as the number of TN increased for additional objects and all the TN were identified correctly. This shows that the CQL-VA operators can detect  a situation correctly, even if other arbitrary situations are present in the video.

\begin{table}[ht]
\centering
\caption { \textmd{\textbf{Robustness of Q3}.  $3437^{+}$ and $3439^{+}$ generated by adding arbitrary videos not containing Q3 situations to 3437 and 3439, respectively. Join, cJoin and CCT join produced same accuracy in this experiment.}}
\begin{tabular}{|p{0.8cm}|p{1cm}|p{0.73cm}|p{0.73cm}|p{0.9cm}|p{0.7cm}|p{0.9cm}|}
\hline
\textbf{Dataset} &  \textbf{$V_{id_1} \bowtie V_{id_2}$ }  & \textbf{$V_{id_1}$ objects}  & \textbf{$V_{id_2}$ objects} & \textbf{GT(vce)}  & \textbf{cJoin} & \textbf{Acc(vce)}\\\hline

\multirow{2}{1 cm}{MavVid \cite{DataSet/MavVid}} & $3437 \bowtie 3439$ & O$_1$,O$_2$, O$_3$,O$_5$ & O$_1$,O$_3$ & (O$_1$,O$_1$), (O$_3$,O$_3$) & TP=1, FP=2, FN=1, \textbf{TN=4} & 62.5\% \\ \cline{2-7}

~ & $3437^{+} \bowtie 3439^{+}$ & O$_1$,O$_2$, O$_3$,O$_5$, O$_6$ & O$_1$, O$_3$,O$_7$ & (O$_1$,O$_1$), (O$_3$,O$_3$) & TP=1, FP=2, FN=1, \textbf{TN=11} & 80\%  \\
 \hline
\end{tabular}
\label{tab:Query3-robustness}
\qquad
\end{table}

\section{Conclusions and Future Work}
\label{sec:conclusions}

\noindent In this paper, we have provided a novel database framework and identified the components for general-purpose video analysis and shown how a few additional operators can detect situations efficiently. We have shown how established and low-risk approaches can be adapted for the new domain of query-based situation monitoring. 
This is only a starting point for establishing the viability of the database framework for content-based video situation monitoring. There are several challenges from here to automated real-time situation monitoring. some of them are:
\begin{packed_itemize}
\item Combining multiple pre-processing outputs for the same video for enhancing extracted contents
\item Complex situations detection including using event detection on video contents
\item New operators (spatial, temporal, object interaction, ...), their semantics, and efficient computation and new window types
\item Optimization of CQL-VA continuous queries
\item Automatic inference of \textit{th} (threshold) value (can benefit from machine learning techniques)
\item Real-time issues: runtime monitoring, scheduling, load shedding, capacity modeling 
\item Supporting dynamic multiple CQ processing and exploit shared sub-computations
\end{packed_itemize}



\scriptsize
\bibliographystyle{IEEEtran}
 \bibliography{./bibliography/itlabPublications,./bibliography/itlabTheses,./bibliography/sp-book,./bibliography/afrl-2017,./bibliography/hafsaResearch, ./bibliography/aleksandric}

\begin{thebibliography}{10}
\providecommand{\url}[1]{#1}
\csname url@samestyle\endcsname
\providecommand{\newblock}{\relax}
\providecommand{\bibinfo}[2]{#2}
\providecommand{\BIBentrySTDinterwordspacing}{\spaceskip=0pt\relax}
\providecommand{\BIBentryALTinterwordstretchfactor}{4}
\providecommand{\BIBentryALTinterwordspacing}{\spaceskip=\fontdimen2\font plus
\BIBentryALTinterwordstretchfactor\fontdimen3\font minus
  \fontdimen4\font\relax}
\providecommand{\BIBforeignlanguage}[2]{{%
\expandafter\ifx\csname l@#1\endcsname\relax
\typeout{** WARNING: IEEEtran.bst: No hyphenation pattern has been}%
\typeout{** loaded for the language `#1'. Using the pattern for}%
\typeout{** the default language instead.}%
\else
\language=\csname l@#1\endcsname
\fi
#2}}
\providecommand{\BIBdecl}{\relax}
\BIBdecl

\bibitem{objectRecognintion/redmon2018yolov3}
J.~Redmon and A.~Farhadi, ``Yolov3: An incremental improvement,'' \emph{arXiv
  preprint arXiv:1804.02767}, 2018.

\bibitem{objectRecognintion/MRCNN}
K.~He, G.~Gkioxari, P.~Doll{\'a}r, and R.~Girshick, ``Mask r-cnn,'' in
  \emph{Proceedings of the IEEE international conference on computer vision},
  2017, pp. 2961--2969.

\bibitem{videoQuerying/yadav2021vidwin}
P.~Yadav, D.~Salwala, and E.~Curry, ``Vid-win: Fast video event matching with
  query-aware windowing at the edge for the internet of multimedia things,''
  \emph{IEEE Internet of Things Journal}, vol.~8, no.~13, pp. 10\,367--10\,389,
  2021.

\bibitem{AmazonKinesis}
J.~Varia, S.~Mathew \emph{et~al.}, ``Overview of amazon web services,''
  \emph{Amazon Web Services}, vol. 105, 2014.

\bibitem{videoQuerying/aref2003video}
W.~Aref, M.~Hammad, A.~C. Catlin, I.~Ilyas, T.~Ghanem, A.~Elmagarmid, and
  M.~Marzouk, ``Video query processing in the vdbms testbed for video database
  research,'' in \emph{Proceedings of the 1st ACM international workshop on
  Multimedia databases}, 2003, pp. 25--32.

\bibitem{vqa/xiong2019visual}
P.~Xiong, H.~Zhan, X.~Wang, B.~Sinha, and Y.~Wu, ``Visual query answering by
  entity-attribute graph matching and reasoning,'' in \emph{Proceedings of the
  IEEE/cvf conference on computer vision and pattern recognition}, 2019, pp.
  8357--8366.

\bibitem{lerner2003aquery}
A.~Lerner and D.~Shasha, ``Aquery: Query language for ordered data,
  optimization techniques, and experiments,'' in \emph{Proceedings of the 29th
  international conference on Very large data bases-Volume 29}.\hskip 1em plus
  0.5em minus 0.4em\relax VLDB Endowment, 2003, pp. 345--356.

\bibitem{arrayDB/baumann2021array}
P.~Baumann, D.~Misev, V.~Merticariu, and B.~P. Huu, ``Array databases:
  concepts, standards, implementations,'' \emph{Journal of Big Data}, vol.~8,
  no.~1, pp. 1--61, 2021.

\bibitem{videoQuerying/kang2017noscope}
D.~Kang, J.~Emmons, F.~Abuzaid, P.~Bailis, and M.~Zaharia, ``Noscope:
  optimizing neural network queries over video at scale,'' \emph{arXiv preprint
  arXiv:1703.02529}, 2017.

\bibitem{videoQuerying/bastani2020miris}
F.~Bastani, S.~He, A.~Balasingam, K.~Gopalakrishnan, M.~Alizadeh,
  H.~Balakrishnan, M.~Cafarella, T.~Kraska, and S.~Madden, ``Miris: Fast object
  track queries in video,'' in \emph{Proceedings of the 2020 ACM SIGMOD
  International Conference on Management of Data}, 2020, pp. 1907--1921.

\bibitem{videoQuerying/chao2020svq}
D.~Chao, N.~Koudas, and I.~Xarchakos, ``Svq++: Querying for object interactions
  in video streams,'' in \emph{Proceedings of the 2020 ACM SIGMOD International
  Conference on Management of Data}, 2020, pp. 2769--2772.

\bibitem{videoQuerying/kang2018blazeit}
D.~Kang, P.~Bailis, and M.~Zaharia, ``Blazeit: optimizing declarative
  aggregation and limit queries for neural network-based video analytics,''
  \emph{arXiv preprint arXiv:1805.01046}, 2018.

\bibitem{videoQuerying/donderler2005bilvideo}
M.~E. D{\"o}nderler, E.~{\c{S}}aykol, U.~Arslan, {\"O}.~Ulusoy, and
  U.~G{\"u}d{\"u}kbay, ``Bilvideo: Design and implementation of a video
  database management system,'' \emph{Multimedia Tools and Applications},
  vol.~27, no.~1, pp. 79--104, 2005.

\bibitem{videoQuerying/svql2015}
C.~Lu, M.~Liu, and Z.~Wu, ``Svql: A sql extended query language for video
  databases,'' \emph{International Journal of Database Theory and Application},
  vol.~8, no.~3, pp. 235--248, 2015.

\bibitem{aved2014informatics}
A.~J. Aved and K.~A. Hua, ``An informatics-based approach to object tracking
  for distributed live video computing,'' \emph{Multimedia Tools and
  Applications}, vol.~68, no.~1, pp. 111--133, 2014.

\bibitem{Babcock:PODS02:Models}
B.~Babcock, S.~Babu, M.~Datar, R.~Motwani, and J.~Widom, ``Models and issues in
  data stream systems.'' in \emph{Proceedings of the Twenty-first ACM
  SIGACT-SIGMOD-SIGART Symposium on Principles of Database Systems}, June 2002,
  pp. 1--16.

\bibitem{DBLP:journals/vldb/ArasuBW06}
A.~Arasu, S.~Babu, and J.~Widom, ``{The CQL continuous query language: semantic
  foundations and query execution},'' \emph{VLDB Journal}, vol.~15, no.~2, pp.
  121--142, 2006.

\bibitem{DBLP:journals/sigmod/GhanemAE06}
T.~M. Ghanem, W.~G. Aref, and A.~K. Elmagarmid, ``{Exploiting predicate-window
  semantics over data streams},'' \emph{SIGMOD Record}, vol.~35, no.~1, pp.
  3--8, 2006.

\bibitem{thein2014apache}
K.~M.~M. Thein, ``Apache kafka: Next generation distributed messaging system,''
  \emph{International Journal of Scientific Engineering and Technology
  Research}, vol.~3, no.~47, pp. 9478--9483, 2014.

\bibitem{Carbone:2017:SMA:3137765.3137777}
\BIBentryALTinterwordspacing
P.~Carbone, S.~Ewen, G.~F\'{o}ra, S.~Haridi, S.~Richter, and K.~Tzoumas,
  ``State management in apache flink\&reg;: Consistent stateful distributed
  stream processing,'' \emph{Proc. VLDB Endow.}, vol.~10, no.~12, pp.
  1718--1729, Aug. 2017. [Online]. Available:
  \url{https://doi.org/10.14778/3137765.3137777}
\BIBentrySTDinterwordspacing

\bibitem{2015TwitterHS}
``Twitter heron: Stream processing at scale,'' in \emph{SIGMOD Conference},
  2015.

\bibitem{VideoRepresentation/yadav2020knowledge}
P.~Yadav, D.~Salwala, D.~P. Das, and E.~Curry, ``Knowledge graph driven
  approach to represent video streams for spatiotemporal event pattern matching
  in complex event processing,'' \emph{International Journal of Semantic
  Computing}, vol.~14, no.~03, pp. 423--455, 2020.

\bibitem{Wojke2017simple}
N.~Wojke, A.~Bewley, and D.~Paulus, ``Simple online and realtime tracking with
  a deep association metric,'' in \emph{2017 IEEE International Conference on
  Image Processing (ICIP)}.\hskip 1em plus 0.5em minus 0.4em\relax IEEE, 2017,
  pp. 3645--3649.

\bibitem{Book/Chakravarthy09}
S.~Chakravarthy and Q.~Jiang, \emph{{Stream Data Management: A Quality of
  Service Perspective}}.\hskip 1em plus 0.5em minus 0.4em\relax Springer, April
  2009.

\bibitem{DEBS/Chakravarthy08}
S.~Chakravarthy and R.~Adaikkalavan, ``{Event and Streams: Harnessing and
  Unleashing Their Synergy},'' in \emph{International Conference on Distributed
  Event-based Systems}, July 2008, pp. 1--12.

\bibitem{ICDT/Jiang07}
Q.~Jiang, R.~Adaikkalavan, and S.~Chakravarthy, ``{MavEStream: Synergistic
  Integration of Stream and Event Processing},'' in \emph{International
  Conference on Digital Communications}, 2007, pp. 29--29.

\bibitem{DataSet/Camnet}
S.~Zhang, E.~Staudt, T.~Faltemier, and A.~K. Roy-Chowdhury, ``A camera network
  tracking (camnet) dataset and performance baseline,'' in \emph{2015 IEEE
  Winter Conference on Applications of Computer Vision}.\hskip 1em plus 0.5em
  minus 0.4em\relax IEEE, 2015, pp. 365--372.

\bibitem{DataSet/MavVid}
``Mavvid dataset prepared in itlab,''
  \url{https://itlab.uta.edu/downloads/datasets/MavVid.zip}.

\bibitem{Implementation/YOLO}
``Yolo source code,'' \url{https://github.com/ZQPei/deep_sort_pytorch}.

\bibitem{Dataset/CVIU_UA-DETRAC}
L.~Wen, D.~Du, Z.~Cai, Z.~Lei, M.~Chang, H.~Qi, J.~Lim, M.~Yang, and S.~Lyu,
  ``{UA-DETRAC:} {A} new benchmark and protocol for multi-object detection and
  tracking,'' \emph{Computer Vision and Image Understanding}, 2020.

\end{thebibliography}


\end{document}